Electron-Phonon Coupling from *Ab Initio* Linear-Response Theory within the *GW* Method: Correlation-Enhanced Interactions and Superconductivity in Ba$_{1-x}$K$_x$BiO$_3$

Zhenglu Li, Gabriel Antonius, Meng Wu, Felipe H. da Jornada, and Steven G. Louie*

*Department of Physics, University of California at Berkeley, California 94720, USA and*

*Materials Sciences Division, Lawrence Berkeley National Laboratory, Berkeley, California 94720, USA*

*Email: sglouie@berkeley.edu

**Abstract**

We present a new first-principles linear-response theory of changes due to perturbations in the quasiparticle self-energy operator within the *GW* method. This approach, named *GW* perturbation theory (*GW*PT), is applied to calculate the electron-phonon (*e*-ph) interactions with the full inclusion of the *GW* non-local, energy-dependent self-energy effects, going beyond density-functional perturbation theory. Avoiding limitations of the frozen-phonon technique, *GW*PT gives access to *e*-ph matrix elements at the *GW* level for *all* phonons and scattering processes, and the computational cost scales *linearly* with the number of phonon modes (wavevectors and branches) investigated. We demonstrate the capabilities of *GW*PT by studying the *e*-ph coupling and superconductivity in Ba$_{0.6}$K$_{0.4}$BiO$_3$. We show that many-electron correlations significantly enhance the *e*-ph interactions for states near the Fermi surface, and explain the observed high superconductivity transition temperature of Ba$_{0.6}$K$_{0.4}$BiO$_3$ as well as its doping dependence.

First-principles calculation of electron-phonon (*e*-ph) coupling [1] is of tremendous interest as it serves as a non-empirical approach to predict and understand a number of phenomena in condensed matter physics and materials physics, such as phonon-mediated superconductivity, electrical and thermal transport, quasiparticle energy renormalization, charge-density wave (CDW), and vibrational features in optical spectra. By formulating a linear-response theory of density functional theory (DFT) [2] to phonon perturbations, density-functional perturbation theory (DFPT) [3–6] has been the prevailing and most efficient *ab initio* method to study the *e*-ph interactions within DFT. The *e*-ph coupling treated in DFPT is only approximate since the DFT orbital eigenvalues are not the true electron (or quasiparticle) energies. This is reflected in that, in general, the Kohn-Sham eigenvalues do not yield accurate band gaps and band widths nor information on lifetimes [7,8]. The exchange-correlation potentials $V_{xc}$ in DFT (such as those in the local-density approximation (LDA) [9] or the generalized gradient approximation (GGA) [10]) can only be at best considered as an approximation to the nonlocal, frequency-dependent self-energy operator $\Sigma$.

The *GW* approximation [7,8,11–13] has proven, for many materials, to be an accurate *ab initio* method in capturing the many-electron correlation effects in the evaluation of the quasiparticle energies. In the *GW* approximation, the self-energy operator $\Sigma$ is expanded in terms of the single-particle Green's function $G$ and the screened Coulomb interaction $W$ to first order, i.e. $\Sigma = iGW$, hence named the *GW* method. By combining frozen-phonon technique with *GW* calculations, previous studies [14–20] have found that many-electron *corrections* to DFT *e*-ph coupling strength are essential to accurately describe a number of phenomena, such as the phonon dispersion in graphene and graphite [14,15], the temperature-dependent band gap in diamond [18], and superconductivity in $Ba_{0.6}K_{0.4}BiO_3$ [17]. However, the frozen-phonon technique is limited to only investigate couplings to phonon wavevectors that are commensurate to a large supercell, which makes it prohibitive to achieve a fine sampling of the Brillouin zone (BZ). More importantly, frozen-phonon calculations can only provide *some* intra-band part of the *e*-ph matrix elements *indirectly* and an overall *e*-ph coupling strength by examining band energy shifts. The *e*-ph matrix elements between different bands and for wavevectors across the full BZ – the essential ingredient of microscopic *e*-ph formulations of many physical phenomena – are *not* available within frozen-phonon methods [1,17,19,20]. The importance of self-energy effects in *e*-ph coupling and the severe limitations of the frozen-phonon *GW* technique thus point to a strong necessity for a linear-response *GW* theory (similar in spirit as DFPT [3-6]) to efficiently and accurately calculate the quasiparticle *e*-ph interactions at the *GW* level [1,17,19,20].

In this Letter, for the first time, we present a first-principles linear-response *GW* method to external perturbations, which we call *GW* perturbation theory (*GW*PT). In this scheme, the first-order change of the self-energy operator to a phonon perturbation $\Delta_{\mathbf{q}\nu}\Sigma$ is constructed from a linear-response calculation, which is performed within a single primitive unit cell for any phonon wavevector $\mathbf{q}$ and phonon branch $\nu$. This method avoids the use of supercells, and the computational cost naturally scales linearly with the number of phonon modes needed. More importantly from a physics view point, it provides the *e*-ph matrix elements at the *GW* level for any pairs of electronic states directly and efficiently, making *GW*PT a desirable *ab initio* method to systematically study *e*-ph interactions including many-electron self-energy effects. We demonstrate the power of the *GW*PT method by studying the *e*-ph coupling and superconductivity in $Ba_{0.6}K_{0.4}BiO_3$ [21-23] as well as other doping concentrations away from $x = 0.4$. We find that the *GW* self-energy renormalizes the DFT-LDA *e*-ph matrix elements *non-uniformly* across the BZ, and enhances the *e*-ph coupling constant $\lambda$ by a factor of 2.4. The *GW*PT-calculated $\lambda = 1.14$ is strong enough to account for the high superconducting transition temperature $T_c$ in $Ba_{0.6}K_{0.4}BiO_3$. We show that the doping dependence in the superconductivity is mainly from a density-of-states (DOS) effect.

Here the *GW*PT method is formulated for phonon perturbations, but it can straightforwardly be generalized to other perturbations such as electric field and strain. We present the theory for crystals with time-reversal symmetry (TRS), and spin indices are omitted for simplicity. The key quantity is the *e*-ph coupling matrix element $g_{mn\nu}(\mathbf{k}, \mathbf{q})$. The *e*-ph matrix element at the *GW* level can be constructed in a similar way that the quasiparticle energy is constructed [8], with the contribution from the *GW* self-energy replacing that from the exchange-correlation functional $V_{xc}(\mathbf{r})$ in DFT; that is,

$$g_{mn\nu}^{GW}(\mathbf{k}, \mathbf{q}) = g_{mn\nu}^{\mathrm{DFT}}(\mathbf{k}, \mathbf{q}) - \langle \psi_{m\mathbf{k}+\mathbf{q}} | \Delta_{\mathbf{q}\nu} V_{xc} | \psi_{n\mathbf{k}} \rangle + \langle \psi_{m\mathbf{k}+\mathbf{q}} | \Delta_{\mathbf{q}\nu} \Sigma | \psi_{n\mathbf{k}} \rangle,$$

(1)

where

$$g_{mn\nu}^{\mathrm{DFT}}(\mathbf{k}, \mathbf{q}) = \langle \psi_{m\mathbf{k}+\mathbf{q}} | \Delta_{\mathbf{q}\nu} V^{\mathrm{KS}} | \psi_{n\mathbf{k}} \rangle$$

(2)

is the *e*-ph matrix element at the DFT level [1]. $V^{\mathrm{KS}}$ is the total Kohn-Sham potential in DFT, and $\psi_{n\mathbf{k}}$ and $\psi_{m\mathbf{k}+\mathbf{q}}$ are the wavefunctions of the initial and final electron states involved in the scattering process, with band indices *n* and *m* at wavevectors **k** and **k**+**q**, respectively. The differential perturbation operator $\Delta_{\mathbf{q}\nu}$ gives the linear change in the quantity it operates on, when the system is perturbed with a phonon mode labeled by $\mathbf{q}\nu$ with the atoms displaced by the zero-point displacement amplitude [1]. The dimensionless operator $\Delta_{\mathbf{q}\nu}$ carries a crystal momentum of **q**, and is explicitly defined as

$$\Delta_{\mathbf{q}\nu} = \sqrt{\frac{\hbar}{2\omega_{\mathbf{q}\nu}}} \sum_{\kappa\alpha} \frac{1}{\sqrt{M_\kappa}} e_{\kappa\alpha,\nu}(\mathbf{q}) \sum_{l}^{N_l} e^{i\mathbf{q}\cdot\mathbf{R}_l} \frac{\partial}{\partial \tau_{\kappa\alpha l}},$$

(3)

where $\alpha = x, y, z$ labels the Cartesian directions, $\kappa$ counts the atoms in the primitive unit cell, $M_\kappa$ is the mass of the $\kappa$-th atom, $e_{\kappa\alpha,\nu}(\mathbf{q})$ is the $\kappa\alpha$-component of the $\nu$-th eigenvector of the dynamical matrix at **q**, $\omega_{\mathbf{q}\nu}$ is the phonon frequency, *l* labels the *l*-th unit cell in the material, and $\mathbf{R}_l$ is the corresponding lattice position vector. In Eq. (3), the partial derivative is taken with respect to the atom coordinate $\tau_{\kappa\alpha l}$ of the $\kappa$-th atom, along the $\alpha$ direction, and in the *l*-th unit cell. DFPT calculates $\Delta_{\mathbf{q}\nu} V^{\mathrm{KS}}$ by self-consistently solving the Sternheimer equation [1,6]. The Bloch wavefunction has the form $\psi_{n\mathbf{k}}(\mathbf{r}) = N_l^{-1/2} e^{i\mathbf{k}\cdot\mathbf{r}} u_{n\mathbf{k}}(\mathbf{r})$ where $u_{n\mathbf{k}}(\mathbf{r})$ is a lattice-periodic function.

Therefore, the calculation of the *e*-ph matrix elements at either DFPT or *GW*PT level is indeed done in a primitive unit cell, and no supercells are needed.

Now we construct the change in the self-energy operator. A change in $\Sigma$ involves changes in *G* and *W*. Here, we use the constant-screening approximation [19] such that $\Delta_{\mathbf{q}\nu}W$ may be neglected compared to $\Delta_{\mathbf{q}\nu}G$ against small perturbations. The validity of this approximation has been verified by using frozen-phonon calculations in a previous study [19] and by our own calculations. It is expected to be generally valid in semiconductors where the charges are bounded in bonds, and in metals with large Fermi surfaces. With this approximation, the change in the self-energy operator in the frequency domain reads,

$$\Delta_{\mathbf{q}\nu}\Sigma(\mathbf{r},\mathbf{r}';\varepsilon) = i\int \frac{d\varepsilon'}{2\pi} e^{-i\delta\varepsilon'} \Delta_{\mathbf{q}\nu}G(\mathbf{r},\mathbf{r}';\varepsilon-\varepsilon')W(\mathbf{r},\mathbf{r}';\varepsilon'),$$

(4)

where $\varepsilon$ and $\varepsilon'$ are energy variables, and $\delta = 0^+$. To construct $\Delta_{\mathbf{q}\nu}G$, we need the first-order change in the wavefunction [6],

$$\Delta_{\mathbf{q}\nu}\psi_{n\mathbf{k}}(\mathbf{r}) = \sum_m \frac{g_{mn\nu}^{\mathrm{DFT}}(\mathbf{k},\mathbf{q})}{\varepsilon_{n\mathbf{k}} - \varepsilon_{m\mathbf{k}+\mathbf{q}}} \psi_{m\mathbf{k}+\mathbf{q}}(\mathbf{r}),$$

(5)

where $\varepsilon_{n\mathbf{k}}$ and $\varepsilon_{m\mathbf{k}+\mathbf{q}}$ are the DFT eigenvalues. Using the knowledge that DFT eigenfunctions well approximate the quasiparticle wavefunctions of most materials [8], the change in the Green's function is written as,

$$\Delta_{\mathbf{q}\nu}G(\mathbf{r},\mathbf{r}';\varepsilon) = \sum_{n\mathbf{k}} \frac{\Delta_{\mathbf{q}\nu}\psi_{n\mathbf{k}}(\mathbf{r})\psi_{n\mathbf{k}}^*(\mathbf{r}') + \psi_{n\mathbf{k}}(\mathbf{r})[\Delta_{-\mathbf{q}\nu}\psi_{n\mathbf{k}}(\mathbf{r}')]^*}{\varepsilon - \varepsilon_{n\mathbf{k}} - i\delta_{n\mathbf{k}}},$$

(6)

where $\delta_{n\mathbf{k}} = 0^+$ for $\varepsilon_{n\mathbf{k}} < \varepsilon_F$ and $\delta_{n\mathbf{k}} = 0^-$ for $\varepsilon_{n\mathbf{k}} > \varepsilon_F$ at zero temperature, and $\varepsilon_F$ is the Fermi energy. In Eq. (6), we have used $\Delta_{\mathbf{q}\nu}\varepsilon_{n\mathbf{k}} = 0$, which is true for $\forall \mathbf{q} \neq 0$ connecting non-degenerate states (see more discussions in Supplemental Materials [24]).

In our implementation of *GW*PT, a plane-wave basis is used. The matrix element of $\Delta_{\mathbf{q}\nu}\Sigma$ now becomes,

$$\langle\psi_{m\mathbf{k}+\mathbf{q}}|\Delta_{\mathbf{q}\nu}\Sigma(\mathbf{r},\mathbf{r}';\varepsilon)|\psi_{n\mathbf{k}}\rangle$$

$$= \frac{i}{2\pi}\sum_{n'}\sum_{\mathbf{p}\mathbf{G}\mathbf{G}'}\Bigg\{\langle\psi_{m\mathbf{k}+\mathbf{q}}|e^{i(\mathbf{p}+\mathbf{G})\cdot\mathbf{r}}|\Delta_{\mathbf{q}\nu}\psi_{n'\mathbf{k}-\mathbf{p}}\rangle\langle\psi_{n'\mathbf{k}-\mathbf{p}}|e^{-i(\mathbf{p}+\mathbf{G}')\cdot\mathbf{r}'}|\psi_{n\mathbf{k}}\rangle \int d\varepsilon' \frac{W_{\mathbf{G}\mathbf{G}'}(\mathbf{p},\varepsilon')e^{-i\delta\varepsilon'}}{\varepsilon - \varepsilon_{n'\mathbf{k}-\mathbf{p}} - i\delta_{n'\mathbf{k}-\mathbf{p}} - \varepsilon'}$$

$$+ \langle\psi_{m\mathbf{k}+\mathbf{q}}|e^{i(\mathbf{p}+\mathbf{G})\cdot\mathbf{r}}|\psi_{n'\mathbf{k}+\mathbf{q}-\mathbf{p}}\rangle\langle\Delta_{-\mathbf{q}\nu}\psi_{n'\mathbf{k}+\mathbf{q}-\mathbf{p}}|e^{-i(\mathbf{p}+\mathbf{G}')\cdot\mathbf{r}'}|\psi_{n\mathbf{k}}\rangle \int d\varepsilon' \frac{W_{\mathbf{G}\mathbf{G}'}(\mathbf{p},\varepsilon')e^{-i\delta\varepsilon'}}{\varepsilon - \varepsilon_{n'\mathbf{k}+\mathbf{q}-\mathbf{p}} - i\delta_{n'\mathbf{k}+\mathbf{q}-\mathbf{p}} - \varepsilon'}\Bigg\},$$

(7)

where $\mathbf{G}$ and $\mathbf{G}'$ are reciprocal lattice vectors, $n'$ and $\mathbf{p}$ are the band index and wavevector for the internal summation, and $W_{\mathbf{G}\mathbf{G}'}(\mathbf{p},\varepsilon')$ is the screened Coulomb interaction. In the construction of *W*, the full dielectric matrix within the random-phase approximation [33] is used. The Hybertsen-Louie generalized plasmon-pole model [8] is employed in this work for the energy convolution of $\varepsilon'$, and we note that the extension to fully frequency-dependent sampling techniques [34] is straightforward. The energy dependence of $\Delta_{\mathbf{q}\nu}\Sigma(\varepsilon)$ is treated with the strategy that every matrix element is evaluated at both $\varepsilon_{n\mathbf{k}}$ and $\varepsilon_{m\mathbf{k}+\mathbf{q}}$, and the average value is taken [35]. Our calculation shows that the energy dependence of the matrix elements is small. Eq. (7) completes Eq. (1) to get $g_{mn\nu}^{GW}(\mathbf{k},\mathbf{q})$.

The above formalism of *GW*PT has been implemented in the BERKELEYGW code [34,36], and is interfaced with the ABINIT code [37] which provides the DFT and DFPT calculations that generate $\Delta_{\mathbf{q}\nu}V_{\text{xc}}(\mathbf{r})$ and $\Delta_{\mathbf{q}\nu}\psi_{n\mathbf{k}}(\mathbf{r})$ (see Supplemental Materials [24]). Spatial symmetries and TRS are used to reduce the phonon **q**-grid [24]. The development of *GW*PT enables access to a lot of new physics where *e*-ph and many-electron interactions are strongly intertwined, especially in correlated materials. Accurate *e*-ph matrix elements and their distribution across BZ and bands calculated using *GW*PT are essential ingredients in the understanding of a number of important phenomena including superconductivity, electrical/thermal transport, electron/phonon lifetimes due to *e*-ph interactions, and temperature-dependent direct/indirect optical absorptions.

We have applied our *GW*PT method (within a one-shot calculation, i.e. $G_0W_0$PT) to study superconductivity in Ba$_{0.6}$K$_{0.4}$BiO$_3$ in its cubic perovskite phase (Fig. 1(a)), which has an experimentally measured superconducting $T_c$ of $30 - 32$ K [21-23]. Previous *ab initio* studies [17,38] show that the *e*-ph coupling calculated within DFT-LDA is too weak to account for such a high $T_c$ in this material, and frozen-phonon *GW* calculations indicate that many-electron self-energy effects may enhance *e*-ph interactions. However, the latter was estimated from a limited study of only a single **q**-point calculation for one electronic state [17].

We first perform standard DFT and DFPT calculations on Ba$_{0.6}$K$_{0.4}$BiO$_3$ using the GGA functional [10]. The calculated Fermi surface shows a regular rounded cubic shape (Fig. 1(b)), and is strongly nested. We verify our *GW*PT method by comparing its results against reference frozen-phonon *GW* results at a selected high symmetry **q**-vector. We focus on the single band (labeled as $n_0$ and highlighted in Fig. 1(c)) crossing $\varepsilon_F$, which is expected to give the dominant contribution to superconductivity. We are interested in $\mathbf{q} = R$, corresponding to a 2×2×2 supercell with atom displacements (see details of frozen-phonon calculation and more verifications in Supplemental Materials [24]). In the frozen-phonon calculation, the energy of the degenerate states at the BZ boundary (*R'* point in Fig. 1(d)) splits linearly with increasing displacement (when it is small enough). The slope in the change in energy with respect to displacement is given by a specific single *e*-ph matrix element that can be fitted from finite-difference frozen-phonon calculations, or *directly* calculated with the linear-response perturbation theory in a primitive unit cell. This type of *e*-ph matrix elements that connect degenerate states is the only one that frozen-phonon *GW* can relatively accurately calculate by making supercells [17], but *GW*PT can access *all* inter- and intra-band *e*-ph matrix elements across the whole BZ with equal and high accuracy. As shown in Fig. 1(e), we find excellent agreement for this matrix element between frozen-phonon DFT and DFPT, and between frozen-phonon *GW* and *GW*PT, nicely verifying our *GW*PT method. Moreover, the DFPT and *GW*PT results are significantly different, illustrating the importance of the quasiparticle self-energy.

To study superconductivity in $Ba_{0.6}K_{0.4}BiO_3$, we calculate the *e*-ph matrix elements that scatter quasiparticle states within the $n_0$ band by performing both DFPT and *GW*PT calculations on an 8×8×8 **k**-grid (full grid) and **q**-grid (35 irreducible **q**-points) [24]. These electronic states are coupled most strongly by phonons in the highest three optical branches [17,38]. As an illustration, we pick out one high-frequency oxygen stretching and breathing optical branch (labeled as $\nu_0$ [24]), and plot the distribution of the strength of the *e*-ph matrix element $|g_{n_0 n_0 \nu_0}(\mathbf{k},\mathbf{q})|$ varying **k** across the BZ for selected **q**-points. Fig. 2(a-c) show the scatterings for $\mathbf{q} = R$ that are mostly relevant to superconductivity in this material. For this important phonon mode, *GW*PT almost uniformly enhances the value of the *e*-ph matrix elements $g$ as compared to DFPT with an enhancement factor of ~1.6. This is because the character of the states on the Fermi surface of $Ba_{0.6}K_{0.4}BiO_3$ is highly isotropic [39]. However, Fig. 2(d-f) (and Fig. S3 [24]) show strong variances in the distribution of the *e*-ph matrix elements and also in the enhancement factor of *GW*PT over DFPT, due to the wavefunction character changing near the $\Gamma$ point of either the initial or final states. These results, for the first time, systematically reveal the complex nature of many-electron renormalization of the *e*-ph interactions, demonstrating the capability and power of *GW*PT.

We evaluate the superconducting $T_c$ of $Ba_{0.6}K_{0.4}BiO_3$ using the McMillan–Allen-Dynes formula [40,41]. The *e*-ph coupling constant $\lambda$ and the characteristic logarithmic-averaged phonon frequency $\omega_{\log}$ [1,40,41] are calculated using the *e*-ph matrix elements that scatter states within the $n_0$ band for all phonon modes, at both the DFPT and *GW*PT level (Table I). The correlation-enhanced *e*-ph coupling constant is directly reflected in the Eliashberg function $\alpha^2 F(\omega)$ by comparing the results from DFPT and *GW*PT in Fig. 3(a). The effective Coulomb parameter $\mu^*$ [1,40,41] is set to a reasonable physical range in Table I. DFPT severely underestimates the superconducting transition temperature, with the calculated $T_c$ in the range of $0.61 - 6.1$ K for $\mu^*$ in the range of $0.18 - 0.08$. However, *GW*PT significantly increases $T_c$ to the range of $28.5 - 44.8$ K for the same range of $\mu^*$ (Table I), in good agreement with the experimentally measured $T_c$ of $30 - 32$ K [21-23]. These results highlight the importance of many-electron correlation effects in *e*-ph interactions [17] that are well captured by the *GW*PT method.

We further study the doping dependence of the superconductivity in $Ba_{1-x}K_xBiO_3$ (superconductivity is observed experimentally for $x > 0.3$) from first principles, calculated using a rigid-band approximation [24]. Fig. 3(b) shows that the superconducting transition temperatures from *GW*PT nicely reproduce the size and shape of the superconducting half dome (however results from DFPT fail significantly) in the phase diagram observed experimentally [21-23]. At doping concentration smaller than $x = 0.3$, the material is in an *insulating* CDW phase with strong structural distortions induced by phonon instability and the nested Fermi surface [17,23,38,42-44]. After $x = 0.4$, an increase in hole doping concentration $x$ suppresses $T_c$, which is mainly due to a reduced DOS with a shrinking Fermi surface. With a reduced Fermi surface, the number of *e*-ph scattering channels decreases, weakening superconductivity (see Supplemental Materials [24] for more analysis). Our *GW*PT results, along with the recent direct experimental observation of isotropic *s*-wave superconducting gap [39], strongly support that superconductivity in $Ba_{1-x}K_xBiO_3$ originates from unusually large $e$-ph interactions, due to many-electron effects.

In summary, we present the theoretical formulation, practical implementation, and application to $Ba_{1-x}K_xBiO_3$ of the newly developed $GW$PT method. $GW$PT is shown to be able to systematically and accurately investigate the rich $e$-ph physics at the $GW$ level, beyond the accessibility of any other existing *ab initio* methods. The capability of $GW$PT demonstrates its great application potential to the study of the rich $e$-ph physics in a wide-range of materials, going beyond DFT.

This work was supported by the Center for Computational Study of Excited-State Phenomena in Energy Materials (C2SEPEM) funded by the U.S. Department of Energy, Office of Basic Energy Sciences under Contract No. DE-AC02-05CH11231 at the Lawrence Berkeley National Laboratory, as part of the Computational Materials Sciences Program, which provided for theory development, code implementation, and calculations, and by the National Science Foundation under Grant No. DMR-1508412, which provided for benchmark studies and tests. Computational resources were provided by the National Energy Research Scientific Computing Center (NERSC), which is supported by the Office of Science of the U.S. Department of Energy under Contract No. DE-AC02-05CH11231, and the Extreme Science and Engineering Discovery Environment (XSEDE), which is supported by National Science Foundation under Grant No. ACI-1053575. The authors thank Y.-H. Chan, T. Cao, C. S. Ong, and H. J. Choi for helpful discussions.

Table I. Calculated $e$-ph coupling constant $\lambda$, logarithmic-averaged phonon frequency $\omega_{\log}$, and superconducting transition temperature $T_c$ (using the McMillan–Allen-Dynes formula) of $Ba_{0.6}K_{0.4}BiO_3$. The effective Coulomb potential parameter $\mu^*$ is set to a reasonable physical range, giving the corresponding range of $T_c$. The experimentally measured $T_c$ is 30 – 32 K [21,22].

|      | $\lambda$ | $\omega_{\log}$ (K) | $\mu^*$      | $T_c$ (K)   |
|------|-----------|---------------------|--------------|-------------|
| DFPT | 0.47      | 488.2               | 0.18 – 0.08  | 0.61 – 6.1  |
| $GW$PT | 1.14    | 491.3               | 0.18 – 0.08  | 28.5 – 44.8 |

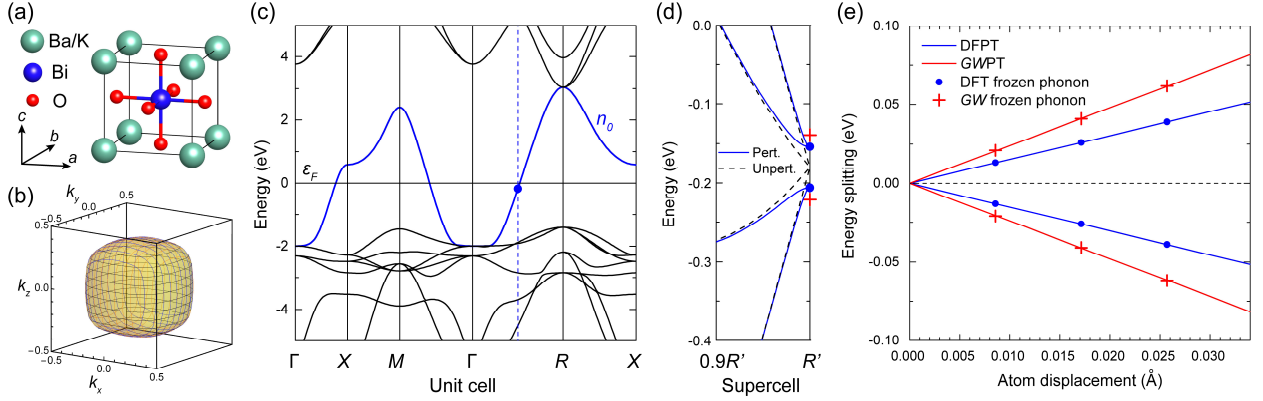

FIG. 1. (a) Crystal structure of $Ba_{0.6}K_{0.4}BiO_3$ in the cubic perovskite phase. (b) Calculated Fermi surface of $Ba_{0.6}K_{0.4}BiO_3$. (c) The DFT-GGA band structure of $Ba_{0.6}K_{0.4}BiO_3$. The band of interest which crosses the $\varepsilon_F$ (set to zero) is highlighted with blue color and labeled as $n_0$. The state at $\mathbf{k} = R/2$ (blue dashed line) indicated by the blue dot has a band energy slightly below $\varepsilon_F$. (d) The DFT band structure of a 2×2×2 supercell. The $R'$ point corresponds to the $\mathbf{k} = R/2$ point at the blue dashed line in (c). The degenerate level indicated by the blue dot in (c) splits upon the oxygen-atom-displacement perturbation (see Supplemental Materials [24]) of 0.0171 Å. The corresponding $GW$ quasiparticle energies are indicated by the red crosses. (e) Comparison of energy splitting-versus-displacement curves between perturbation theory and direct frozen-phonon (finite-difference) calculations.

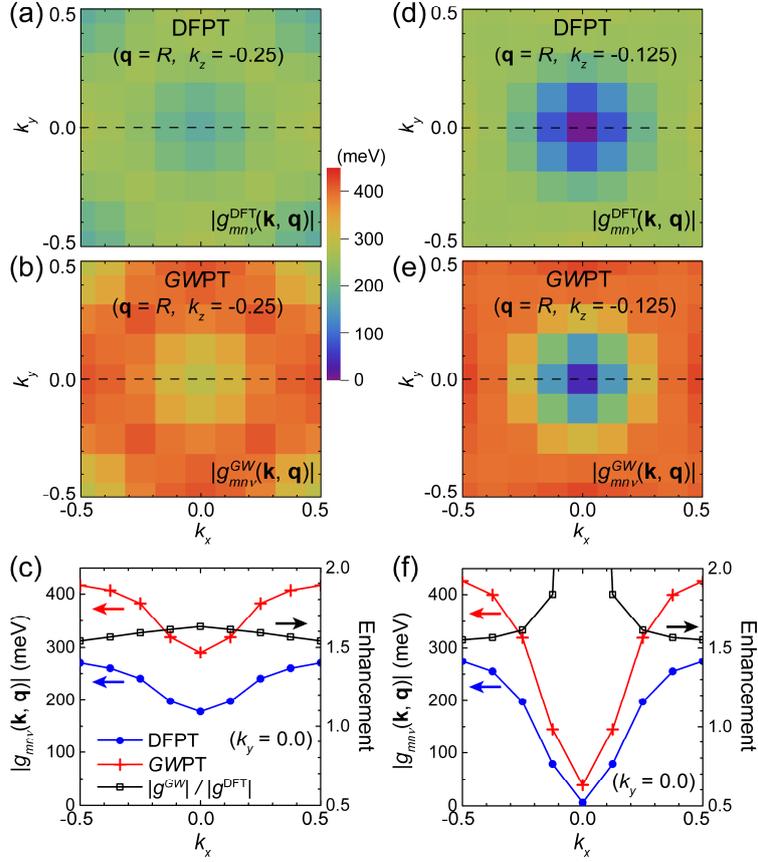

FIG. 2. Distribution of the absolute value of $e$-ph matrix elements $\left|g_{n_0 n_0 \nu_0}(\mathbf{k}, \mathbf{q} = R)\right|$ at (a) DFPT and (b) $GW$PT level for $Ba_{0.6}K_{0.4}BiO_3$, across the $k_x - k_y$ plane at fixed $k_z = -0.25$ of the BZ (reduced coordinates). Calculations are performed on 8×8×8 **k**-grid for each **q**-point. (c) Line profile of (a,b) with $k_y = 0.0$, and the path is indicated by the dashed line in (a,b). The enhancement factor of $|g^{GW}|/|g^{\mathrm{DFT}}|$ is also plotted. (d-f) Similar to (a-c), but with $\mathbf{q} = R$ in the $k_z = -0.125$ plane.

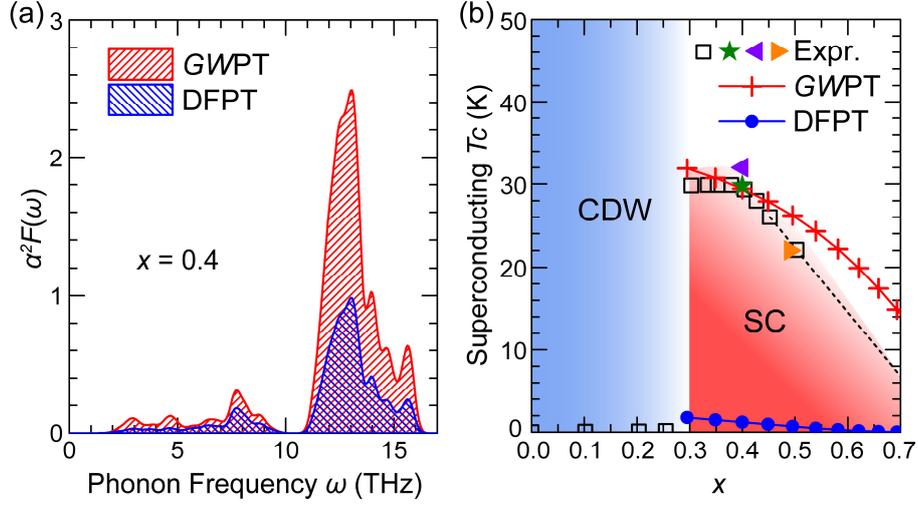

FIG. 3. (a) Eliashberg function $\alpha^2 F(\omega)$ calculated for $Ba_{0.6}K_{0.4}BiO_3$ (i.e. $x = 0.4$) with e-ph matrix elements from *GW*PT and DFPT. (b) Phase diagram of $Ba_{1-x}K_xBiO_3$. Superconducting $T_c$ (red crosses for *GW*PT and blue dots for DFTP) is calculated with $\mu^* = 0.16$. Experimental data are plotted as squares [23], star [21], left-pointing triangle [22], and right-pointing triangle [39]. Black dashed line represents the extrapolation of experimental data into doping range of $x > 0.5$, which is hard to access experimentally. From the superconducting (SC) phase towards undoped parent composition, for $x < 0.3$, the system undergoes a structural phase transition into the non-superconducting CDW phase.

Supplemental Material for "Electron-Phonon Coupling from *Ab Initio* Linear-Response Theory within the *GW* Method: Correlation-Enhanced Interactions and Superconductivity in Ba$_{1-x}$K$_x$BiO$_3$"

Zhenglu Li, Gabriel Antonius, Meng Wu, Felipe H. da Jornada, and Steven G. Louie*

*Department of Physics, University of California at Berkeley, California 94720, USA and*

*Materials Sciences Division, Lawrence Berkeley National Laboratory, Berkeley, California 94720, USA*

*Email: sglouie@berkeley.edu

## I. Computational details

In the construction of the first-order change in the wavefunction $\Delta_{\mathbf{q}\nu}\psi_{n\mathbf{k}}$ in Eq. (5) in the main text, the summation over states avoids numerically unstable effects of degenerate subset of $\varepsilon_{n\mathbf{k}}$ by using a small non-zero imaginary part added to the denominator [1]. Our tests show that this treatment gives negligible effects (this imaginary part can be smaller than the smearing factor used in the density of states calculations, and should not be unphysically large), and is well justified by the excellent agreement between the frozen-phonon *GW* and the *GW*PT results.

In Eq. (6) in the main text, we have neglected the effects induced by the change in $\varepsilon_{n\mathbf{k}}$ due to the phonon perturbations (i.e., using $\Delta_{\mathbf{q}\nu}\varepsilon_{n\mathbf{k}} = 0$). For $\forall \mathbf{q} \neq 0$, any such effects would be from higher-order terms, except for degenerate states connected by a phonon wavevector, which occur in a vanishingly small region of the BZ. In cases of strongly nested systems and low dimensions, such an approximation should be treated carefully, with, e.g., a dense grid for convergence. The $\mathbf{q} = 0$ point is treated at the same footing as other wavevectors in the numerical calculation, since it represents a certain region in the BZ associated with a weight.

In our *GW*PT approach, spatial symmetries and time-reversal symmetry (TRS) are used to reduce the number of phonon **q**-points. The *e*-ph matrix elements on the full uniform **k**-grid are calculated for each irreducible **q**-point. Special care is taken when generating $\Delta_{-\mathbf{q}\nu}\psi_{n\mathbf{k}}$ using TRS. Because $g_{mn\nu}(\mathbf{k}, \mathbf{q})$ is gauge-dependent, to guarantee a consistent gauge throughout the calculation, in the construction of the Green's function, the associated wavefunction and first-order change in the wavefunction must be generated from TRS simultaneously. Therefore, the following substitution is used in Eq. (7) in the main text,

$$|\psi_{n\mathbf{k}}\rangle\langle\Delta_{-\mathbf{q}\nu}\psi_{n\mathbf{k}}| = |\hat{T}\psi_{n-\mathbf{k}}\rangle\langle\hat{T}(\Delta_{\mathbf{q}\nu}\psi_{n-\mathbf{k}})|,$$

(S1)

where $\hat{T}$ is the time-reversal operator. The linear scaling of computational cost in the number of phonon modes and the use of symmetries make *GW*PT very practical in accurately studying the *e*-ph interactions in the presence of many-electron correlation effects. The number of *e*-ph matrix elements to be calculated varies for different physical quantities of interest. For example, for superconductivity, only the *e*-ph matrix elements near the Fermi level are needed.

In this work, the DFT and DFPT calculations of Ba$_{0.6}$K$_{0.4}$BiO$_3$ are performed using ABINIT code [2]. We simulate the potassium (K) doping effects by removing electrons and adding a compensating background charge. The generalized gradient approximation [3] and norm-conserving pseudopotentials [4, 5] are used, with a plane-wave cut-off of 100 Ry. The fully relaxed lattice constant of 4.268 Å is adopted, which is close to the experimental value of 4.283 Å [6]. In the *GW* and *GW*PT calculations using the BERKELEYGW code [7], we take a 15 Ry cut-off for the screened Coulomb interaction, and include 40 empty bands in the construction of the dielectric matrix and the self-energy operator. In metallic systems, the intra-band contributions near the Fermi level are dominant, therefore a relatively small number of empty bands ensures convergence. In the construction of $\Delta_{\mathbf{q}\nu}\Sigma$ operator, the use of Kohn-Sham eigenvalues, DFPT *e*-ph matrix elements, and the unperturbed and first-order wavefunctions from DFT and DFPT, respectively, makes the current calculations at the level of one-shot $G_0W_0$PT.

We set up the frozen-phonon calculations by displacing a single oxygen atom along the normal of the cubic face (that it centers on) as a finite-difference perturbation. This displacement is modulated from one primitive unit cell to another according to the phonon wavevector **q** of interest. We are in particular interested in the **q** = *R* (TRS-equivalent to **q** = −*R*) phonon perturbation, which is (nearly) a nesting vector of the Fermi surface and scatters carriers between the states **k** = −*R*/2 and **k** = *R*/2. Inducing the atom-displacement perturbation in the 2×2×2 supercell splits the degenerate level at the supercell zone boundary (*R'* point in Fig. 1(d) in the main text). The excellent agreement between the frozen-phonon *GW* and *GW*PT shown in Fig. 1(e) in the main text, along with more benchmark calculations we have performed for intrinsic and doped semiconductors [9] (also see Section II), nicely verified our *GW*PT method.

In the frozen-phonon *GW* calculations, the supercell wavefunctions with displacements are used to construct the Σ operator, and the equilibrium wavefunctions are used for the evaluation of the quasiparticle energy. There is a subtlety here – that is, a straightforward/conventional frozen-phonon calculation, such as those in DFT, uses the same wavefunctions with displacements at all time. This is not an issue for the comparison between frozen-phonon DFT and DFPT, because both theories self-consistently update the wavefunctions. However, both the $G_0W_0$ method and $G_0W_0$PT method are not self-consistent theories (A full self-consistency within *GW* theory involves many theoretical and technical challenges and is beyond the scope of this work.), and this leads to some degrees of freedom to build the theory. We use the unperturbed initial and final states wavefunctions to evaluate the first-order change in self-energy operator $\Delta_{\mathbf{q}\nu}\Sigma$ in the *GW*PT theory (as presented in the main text), and this is consistent with the Hellmann-Feynman theorem and the formalism of DFPT. The frozen-phonon *GW* calculations are prepared and performed with the same idea in the supercells. In this way, we benchmark our implementation of the *GW*PT method against the finite-difference frozen-phonon calculations, and see excellent agreement (Fig. 1(e) in the main text and Fig. S1(b)). Note that the assumption that the electrons respond instantaneously to the motion of the ions is used in both DFPT and *GW*PT, therefore the comparison between the perturbation theory and the frozen-phonon method is meaningful.

We further provide an example of comparison of the computation time (defined as total time spent multiplied by number of central processing units used on Intel Knights Landing architecture) between *GW*PT and frozen-phonon *GW*. For the **q** = (0.5, 0.5, 0.5) results plotted in Fig. 1(e) in the main text, the *GW*PT calculation performed in a primitive unit cell takes ~ $10^4$ seconds, and the frozen-phonon *GW* calculation performed in a 2×2×2 supercell takes ~ $10^6$ seconds. The efficiency is differed by two orders of magnitude for this relatively small supercell. For a fine sampling of the **q**-grid, the frozen-phonon *GW* technique becomes prohibitive, whereas *GW*PT has the capability to access the *e*-ph physics systematically and efficiently.

In the main text, we report the results of both DFPT and *GW*PT calculations on 8×8×8 **k**- and **q**-grids (phonon **q**-grid is symmetry reduced). We check the convergence of **k**-grid up to 16×16×16 with DFPT and list the results in Table SI. The superconductivity properties are converged at 8×8×8 **k**-grid, because of the highly isotropic Fermi surface of $Ba_{0.6}K_{0.4}BiO_3$ [10]. The convergence of phonon **q**-grid is usually faster than that of the **k**-grid; therefore, we use the 8×8×8 **q**-grid. In the future, *GW*PT can further be combined with Wannier interpolation techniques [11] to achieve an efficient yet accurate full description of the *e*-ph coupling properties.

## II. Verification example – diamond

We provide another verification example of diamond by comparing frozen-phonon results and perturbation theory calculations, as plotted in Fig. S1. We choose a phonon wavevector **q** = *L*, which corresponds to a 2×1×1 supercell. The state of interest has quadruple degeneracy at the supercell BZ boundary, and will split upon the atom displacement (see Fig. S1). We find excellent agreement between frozen-phonon *GW* and *GW*PT, nicely verifying our *GW*PT method.

## III. Phonon spectrum

The phonon band structure directly obtained with DFPT calculations shows imaginary frequencies at *R* and *M* points, leading to a structural phase transition involving the oxygen octahedron tilting [6, 12-14] at low temperature. Such an instability is identified to be physical both theoretically [13] and experimentally [14]. We calculate a finite-temperature phonon spectrum to get rid of the imaginary frequencies [15] so as to proceed with the *e*-ph calculations. First, a molecular dynamics (MD) sampling at 600 K of a 2×2×2 supercell is performed using the VASP code [16], with the projector augmented wave pseudopotentials [17]. Then 50 randomly selected structures near the MD equilibrium are recalculated using the ABINIT code [2] so that the electronic properties are consistent throughout this work. The calculated forces are then fitted with harmonic potentials to remove the instability, using the ALAMODE code [18]. Finally the inter-atomic force constants and therefore the dynamical matrices and the phonon spectrum are calculated with the fitted potentials.

Our calculated phonon spectrum is in good agreement with neutron diffraction experiment [14] (Fig. S2), except that the frequencies of the oxygen breathing and stretching branch (highlighted in Fig. S2 and we label its mode index as $\nu_0$) are somewhat underestimated. The phonon branches that couple the most to the electronic states near the Fermi level in $Ba_{0.6}K_{0.4}BiO_3$ are the highest three optical branches, as is evidenced by the dominant spectral weight within the range of 11 – 16 THz in the Eliashberg function $\alpha^2 F(\omega)$ (see Fig. 3(a) in the main text).

We note that our calculated *e*-ph coupling constant $\lambda = 0.47$ with DFPT is higher than previous *ab initio* results $\lambda = 0.33$ [12] and $\lambda = 0.29$ [13], and this is largely due to that the previous calculations resulted in higher phonon frequencies (bandwidth ~18 THz) than our calculation (bandwidth ~16 THz, which agrees better with the experiment). Furthermore, the frequency of the strongly e-ph coupled $\nu_0$ branch is around 13 THz from our calculated $\alpha^2F(\omega)$ function (see Fig. 3 in the main text), whereas in Refs. [12, 13] the dominant peak is localized around 17 THz. The overestimation of phonon frequency will lead to an underestimation of $\lambda$, considering which previous calculations [12,13] and our calculated $\lambda$ agree nicely. Nevertheless, the calculated $T_c$ at DFPT level is still too low compared with experiments. However, according to our GWPT calculations, the absolute values of e-ph matrix elements $|g|$ are increased with an overall enhancement factor ~1.6 near the Fermi surface (see main text). As is evidenced in the high-frequency range of the $\alpha^2F(\omega)$ function (see Fig. 3(a) in the main text), the e-ph coupling constant $\lambda$ is largely enhanced by a factor ~ 2.4 (coming from $|g|^2$) in GWPT compared with DFPT. Consequently, the enhancement of $T_c$ due to the many-electron correlation effects is quite significant.

In the calculation of superconductivity, the DFT band structure is used, as our calculated GW band structure remains quite similar near the Fermi level. The significant enhancement of e-ph coupling is dominantly due to the many-electron renormalization of the e-ph matrix elements.

**IV. Distribution of *e*-ph matrix elements in BZ**

We show further results of the distribution of e-ph matrix elements in BZ, in Fig. S3. The combined results from Fig. 2 in the main text and Fig. S3 show that in general, e-ph matrix elements and the many-electron renormalization distributions are quite non-uniform in both the electron and the phonon BZ. For systems with complex Fermi surfaces, and for calculations of many important quantities such as e-ph self-energy, the detailed renormalizations of all e-ph matrix element are needed from GWPT to accurately describe the e-ph physics.

**V. Doping dependence**

In the study of the doping-dependent superconductivity in Ba$_{1-x}$K$_x$BiO$_3$, we adopt the rigid-band approximation (band energy, phonon spectrum, and e-ph matrix elements are taken from *ab initio* calculations at $x = 0.4$), which is usually a good approximation if the doping range of study is not too large so that the system remains similar. By varying the Fermi level for different doping concentrations, we recalculate the Eliashberg function $\alpha^2F(\omega)$, the e-ph coupling constant $\lambda$, and the superconducting $T_c$ (using $\mu^* = 0.16$).

The experimental optimal doping (i.e. with the highest $T_c$) is near $x = 0.4$. Experimentally [6], increased hole doping (i.e., increased $x$) suppresses $T_c$. Our calculated GWPT results nicely reproduce this trend and agrees with experiment [6], as shown in Fig. 3(b) in the main text. This suppression behavior in the over-doped region is understood mainly from a DOS effect. By looking at the band structure (Fig. 1(c) in the main text), an increasing hole doping level leads to smaller DOS (Fig. S4) from a shrinking Fermi surface. Therefore, the number of allowed scatterings on Fermi surface is reduced, leading to the suppressed $T_c$. In Fig. S4, it is clear that the e-ph coupling constant $\lambda$ (from both GWPT and DFPT) follows the trend of DOS as a function of doping $x$. Note that DFPT fails to explain the strength of the superconductivity across the whole range of doping, whereas GWPT reproduces the experimental data [6] well. Beyond the rigid-band approximation, other factors involving modifications in crystal structures, quasiparticle bands, phonon spectrum, and e-ph matrix elements may have further small influences on the doping dependence. However, within $0.3 < x < 0.5$ ($x > 0.5$ is very hard to access experimentally [6]), the rigid-band approximation based on *GW*PT calculations captures the dominant DOS effects, and provides reliable doping dependence in both magnitude and trend.

Table SI. Convergence of **k**-grid for superconductivity properties at DFPT level. The *e*-ph coupling constant $\lambda$, logarithmic-averaged phonon frequency $\omega_{\log}$, and the superconducting transition temperature $T_c$ are calculated using DFPT on different **k**-grids, but with the same 8×8×8 phonon **q**-grid. The effective Coulomb potential parameter $\mu^*$ is set to the same indicated range.

| DFPT **k**-grid | $\lambda$ | $\omega_{\log}$ (K) | $\mu^*$ | $T_c$ (K) |
|---|---|---|---|---|
| 8×8×8 | 0.47 | 488.2 | 0.18 – 0.08 | 0.61 – 6.1 |
| 16×16×16 | 0.51 | 486.4 | 0.18 – 0.08 | 1.2 – 8.2 |

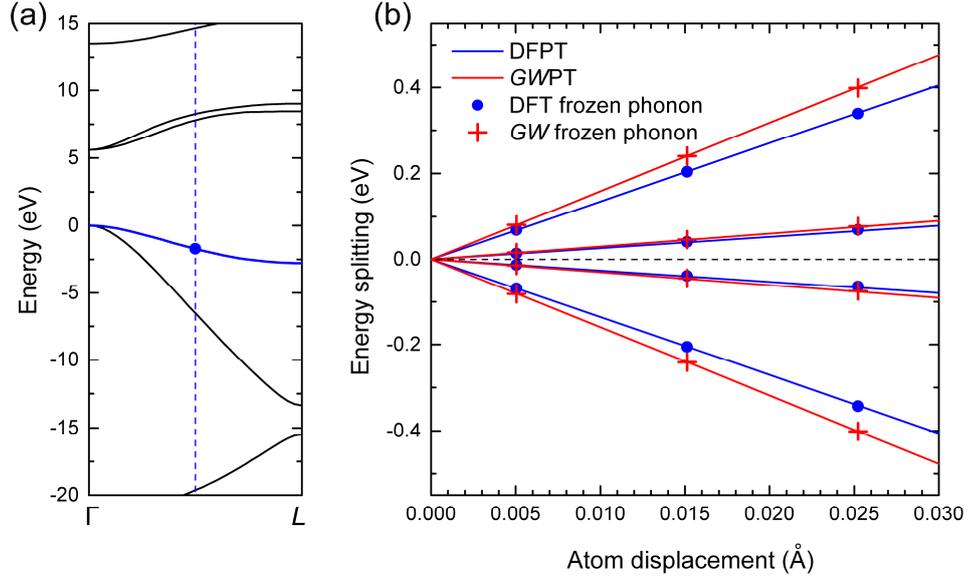

FIG. S1. (a) DFT band structure of diamond, where the valence band maximum is set to zero energy. The high symmetry points are $\Gamma = (0.0, 0.0, 0.0)$ and $L = (0.5, 0.0, 0.0)$. We consider a phonon wavevector **q** = $L$, which corresponds to a 2×1×1 supercell, folding the BZ at the blue dashed line. The highlighted blue band is doubly degenerate, therefore the state of interest (at **k** = $L/2$ in the primitive unit cell BZ, indicated by the blue dot) will have four degenerate states at the supercell BZ boundary. After applying an atom displacement (moving one atom along the first lattice vector), the four states will split. (b) Comparison of energy splitting-versus-displacement curves between perturbation theory and direct frozen-phonon (finite-difference) calculations.

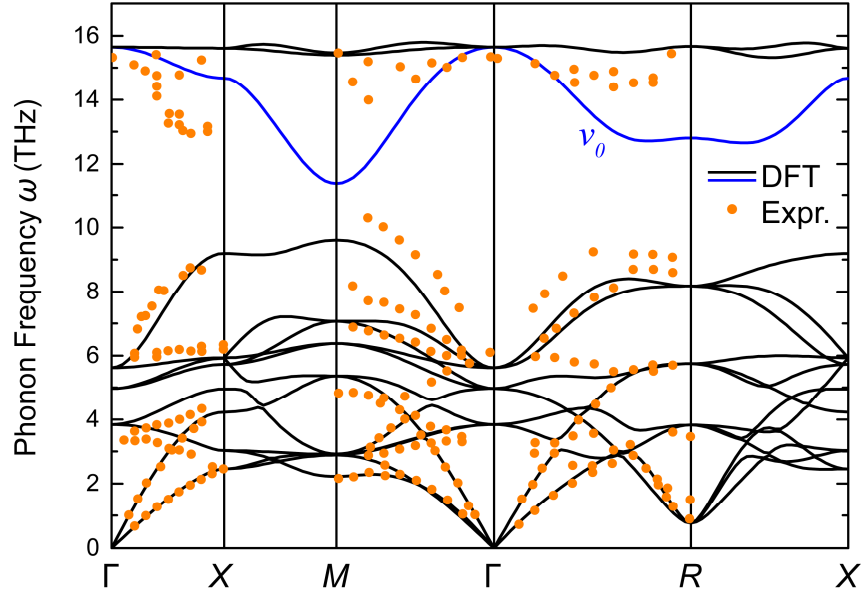

FIG. S2. Phonon band structure of $Ba_{0.6}K_{0.4}BiO_3$ calculated at the DFT level. The high symmetry points are $\Gamma = (0.0, 0.0, 0.0)$, $X = (0.5, 0.0, 0.0)$, $M = (0.5, 0.5, 0.0)$, and $R = (0.5, 0.5, 0.5)$ in units of primitive reciprocal lattice vectors. The dots are data adopted from neutron diffraction experiments [14]. The oxygen breathing and stretching branch is highlighted with blue color and labeled as $\nu_0$.

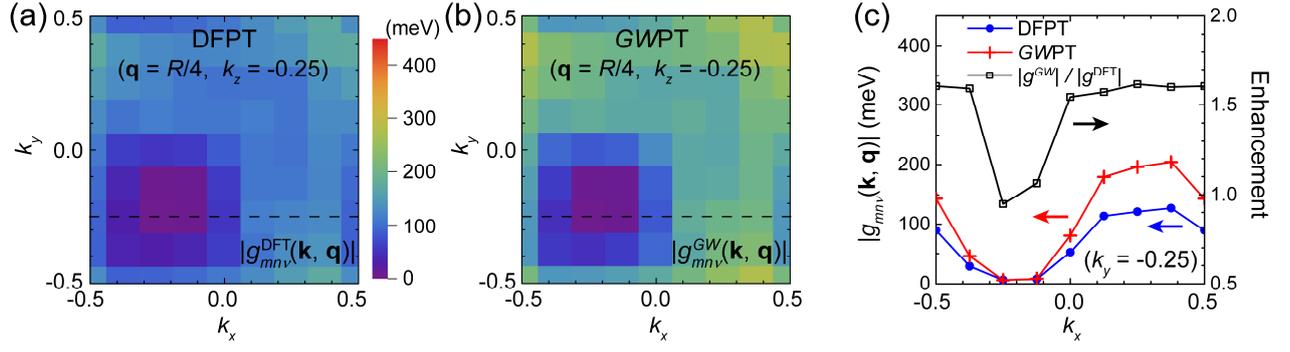

FIG. S3. Distribution of the absolute value of the $e$-ph matrix elements $\left|g_{n_0 n_0 \nu_0}(\mathbf{k},\mathbf{q})\right|$ and the line-profile analysis. (a-c) Similar to Fig. 2(a-c) in the main text, but with $\mathbf{q} = R/4$ in the $k_z = -0.25$ plane.

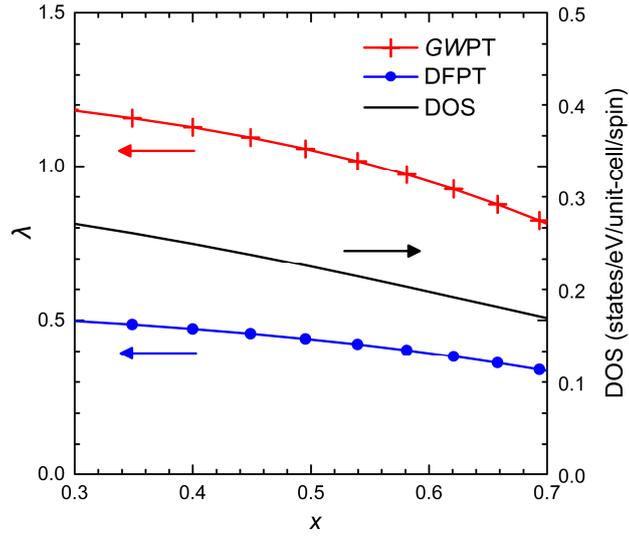

FIG. S4. Doping-dependent *e*-ph coupling constant $\lambda$ from *GW*PT and DFPT, and density of states (DOS) at Fermi level.